\documentclass[aps,preprint,superscriptaddress,amsmath,amssymb]{revtex4-1}

\usepackage[left=2cm,right=2cm,top=2cm,bottom=2cm]{geometry}
\usepackage{amsmath}                 

\usepackage[parfill]{parskip}    
\usepackage{graphicx}
\usepackage{amssymb}
\usepackage{epstopdf}
\usepackage{float}
\usepackage[colorlinks, urlcolor = red]{hyperref}
\usepackage{xcolor}

\usepackage{lipsum}
\begin{document}

\title{A modified galactose network model with implications for growth}
\author{Michele Monti}
\affiliation{ FOM Institute AMOLF, 
Science Park 104
1098 XG Amsterdam,
The Netherlands}
\author{Marta R. A. Matos}
\affiliation{The Novo Nordisk Foundation Center for Biosustainability, Technical University of Denmark, Kogle all\'e 6, 2970 H{\o}rsholm, Denmark}
\author{Jeong-Mo Choi}
\affiliation{Department of Chemistry and Chemical Biology, Harvard University,
12 Oxford Street, Cambridge MA 02138 USA}
\author{Michael S. Ferry}
\affiliation{BioCircuits Institute, University of California, San Diego, 9500 Gilman Drive, La Jolla, CA 92093-0328 USA}
\author{Bart{\l}omiej Borek}
\affiliation{BioCircuits Institute, University of California, San Diego, 9500 Gilman Drive, La Jolla, CA 92093-0328 USA}
\affiliation{baborek@ucsd.edu}

\begin{abstract}
The yeast galactose network has provided many insights into how
eukaryotic gene circuits regulate metabolic function. However, there is currently no consensus model of the network that incorporates protein dilution due to cellular growth. We address this by adapting a well-known model and having it account for growth benefit and burden due to expression of the network proteins. Modifying the model to incorporate galactose transport and basal Gal1p production allows us to better reproduce experimental observations. Incorporating the growth rate effect demonstrates how the native network can optimize growth in
different galactose environments. These findings advance our quantitative
understanding of this gene network, and implement a general approach for
analysing the balance between growth costs and benefits in a range of
metabolic control networks.
\end{abstract}

\maketitle
\section*{Introduction}

The galactose gene (GAL) network in \textit{Saccharomyces cerevisiae} is a model system for the study of transcriptional regulation by galactose and other sugars. The network is composed of: 1) regulatory proteins, such as the transcriptional repressor Gal80p, activator Gal4p, and inducer Gal3p; 2) transporter, Gal2p that imports galactose into the cytoplasm; and 3) the enzymes that metabolize galactose, such as Gal1p, Gal7p, and Gal10p, which convert galactose into glucose-6-phosphate through the Leloir pathway. Gal4p is responsible for activating the expression of all GAL genes except GAL4. However, in the presence of non-inducing and non-repressing carbon sources (such as raffinose) the repressor Gal80p binds to Gal4p, inhibiting the transcription of  the metabolic enzymes and decreasing their protein abundance (the OFF state). When galactose enters the cell it binds to Gal3p which sequesters Gal80p, freeing Gal4p to activate $P_{GAL}$ promoters increasing the expression of metabolic enzymes (ON state) (Fig. \ref{fig:fig1}). Gal1p also has a regulatory role by binding Gal80p, although with weaker affinity than Gal3p \cite{Lavy2015}.

Several quantitative models on the galactose network are available \cite{Atauri2004, Acar2005, Orrell2006, Ramsey2006, Bennett2008, Acar2010, Venturelli2012,cosentino2012, Pannala2012, Apostu2012, Venturelli2015, Peng2015,Nguyen-Huu2015}, but there is no consensus model. Venturelli \textit{et al} \cite{Venturelli2012} combined raffinose to galactose induction experiments and mathematical modeling to demonstrate that the galactose network shows bimodality in protein expression (with a subpopulation of cells persisting in the OFF or ON state) due to a bistability in the model. In contrast, other studies \cite{Ferry2010,Ramsey2006,Acar2005} found that the wild type (WT) strain may not produce monostable responses, but that engineering the strain without the native $P_{GAL80}$ promoter (thus removing feedback from Gal4p to Gal80p) results in bistable responses to galactose. 

During the 9\textsuperscript{th} q-bio Summer School, we worked on a project with the goal of fitting a mathematical model to experiments \cite{Ferry2010} and investigate the effects of adding cell growth. We chose the Venturelli model \cite{Venturelli2012} because of its focus on reproducing the response to galactose OFF/ON transition in the absence of glucose, and its relative simplicity compared to \cite{Atauri2004, Ramsey2006,Pannala2012,Apostu2012}. Here, we present an updated version of the model, and show how the incorporation of growth effects can help optimize metabolic enzyme expression and growth rate for various galactose conditions. 

\section*{Model}

To reproduce some of the galactose induction experiments\cite{Ferry2010} we used all equations and parameters from \cite{Venturelli2012}, except for two modifications: 1) introducing a basal Gal1p production rate, $\alpha^0_{G1}$, to account for Gal1p induction fold changes measured in the experiments; 2) substituting the galactose activation function, $\alpha(Gal)$, with a saturating function, $\alpha_s$ of the form:

\begin{equation}\label{eq:transp}
\alpha_s = C \frac{s}{s+ K_M},
\end{equation}
where $s$ indicates the concentration of galactose outside the cell, $C$ is the maximal transport rate, proportional to the amount of membrane transporters like Gal2p, and $K_M$ is the Michaelis constant of the facilitated diffusion reaction \cite{Barnett1997} (its value taken from \cite{Kasahara11022000}). 

For the WT strain the differential equations for the protein concentrations are:

\begin{equation}\label{mod}
\begin{aligned}
\frac{d[G1]}{dt} &= \alpha^0_{G1} + \alpha_s\epsilon + \alpha_{G1} \frac{[G4]^{n_1}}{[G4]^{n_1} + K_{G1}^{n_1}} + \omega [G1][G80] -\gamma_{G1} [G1]\\
\frac{d[G3]}{dt} &= \alpha_s + \alpha_{G3} \frac{[G4]^{n_3}}{[G4]^{n_3} + K_{G3}^{n_3}} + \delta [G3] [G80] -\gamma_{G3} [G3]\\
\frac{d[G4]}{dt} &= \alpha_{G4} + \beta [G4] [G80] -\gamma_{G4} [G4]\\
\frac{d[G80]}{dt} &= \alpha^0_{G80} + \alpha_{G80} \frac{[G4]^{n_{80}}}{[G4]^{n_{80}} + K_{G80}^{n_{80}}} + \omega [G1] [G80] + \beta[G4][G80] + \delta[G3][G80] -\gamma_{G80} [G80]\\
\end{aligned}
\end{equation}

Where:
\begin{equation}
\begin{aligned}
\omega &= \frac{k_{r81}k_{f81}}{k_{r81}+\gamma_{C81}}-k_{f81}\\
\delta &= \frac{k_{r83}k_{f83}}{k_{r83}+\gamma_{C83}}-k_{f83}\\
\beta &= \frac{k_{r84}k_{f84}}{k_{r84}+\gamma_{C84}}-k_{f84}\\
\end{aligned}
\end{equation}

For the $\Delta P_{GAL80}$ strain, the production terms for [G80] in equation \ref{mod} are replaced with constant synthesis $\alpha^\prime_{G80}$ such that:

\begin{equation}\label{mod2}
\frac{d[G80]}{dt} = \alpha^\prime_{G80} + \omega [G1] [G80] + \beta[G4][G80] + \delta[G3][G80] -\gamma_{G80} [G80]
\end{equation}

\begin{figure}[t]
    \includegraphics[scale=0.34]{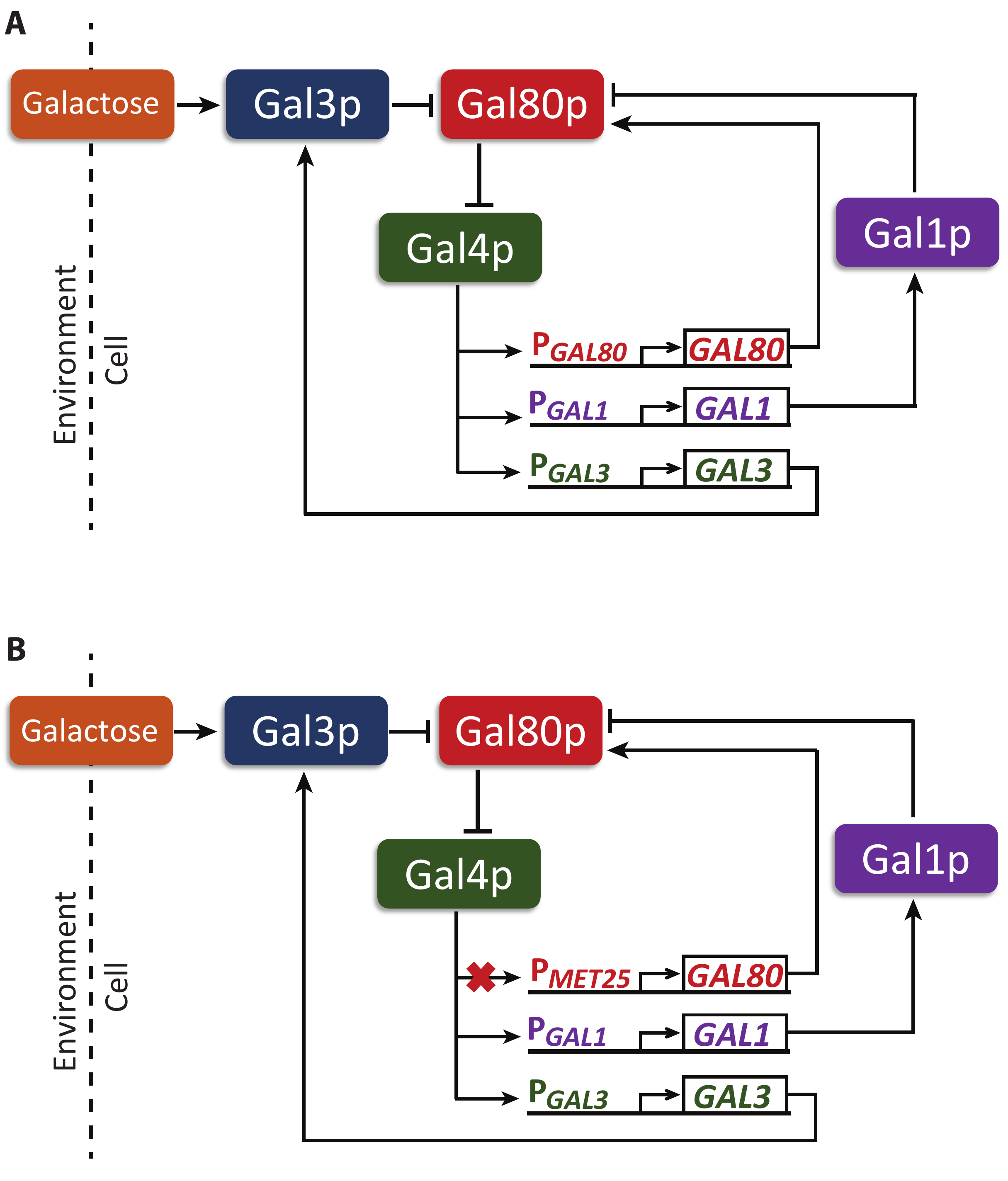}
    \caption{Galactose network in \textit{Saccharomyces cerevisiae} for \textbf{A} the wild type strain and \textbf{B} the $\Delta P_{GAL80}$ strain, where the GAL80 promoter is substituted by a MET25 promoter.}
    \label{fig:fig1}
\end{figure}

Please refer to the Supplementary Section for the meaning and value of each parameter, $p$. The schematic diagrams for the two models are shown in Figure \ref{fig:fig1}.

\section*{Methods}

\subsection*{Strain generation}

{\it S. cerevisiae} K699 was used as the base strain for the experiments. The ''wild type'' strain used (WT, MFSC73) had Gal1p replaced with a Gal1-yCCFP fusion. yCCFP is a yeast enhanced version of cyan fluorescent protein, obtained from the O'Shea Lab (\cite{Raser2004}). Fusion sequences were constructed by double fusion PCR \cite{Wach1996}. The fusion product was combined with ~500bp sequences homologous to the target gene, and transformed with hygromicyn B resistance (HygR). For the negative feedback knockout strain ($\Delta P_{GAL80}$, MFSC121) the native GAL80 promoter, $P_{GAL80}$, was replaced with the methionine repressible $P_{MET25}$ promoter (integrated with G418 resistance). 500$\mu$M of methionine resulted in intermediate expression of Gal80p (see \cite{Ferry2010} p.125). 

\subsection*{Galactose induction experiments}

For galactose induction experiments, the strains were grown at 30$^\circ$~C in synthetic complete medium with 2\% raffinose. They were diluted to OD$_{600}$=0.01 and induced in SC with 2\% raffinose and a given galactose concentration for 8 hours. Fluorescence measurements were taken using a BD LSRII flow cytometer. Signal acquisition and processing is detailed in \cite{Ferry2010} (pp. 106-110).

\subsection*{Model fitting}

For galactose induction fitting, the model curves were generated by integrating trajectories of Eqs. (\ref{mod}),(\ref{mod2}) from an ensemble of initial conditions to steady-state. The mean, $m_m$ and standard deviation $\sigma_m$ of the G1 fold change at steady state were compared to those in the experiments ($m_e, \sigma_e$), using the error function, $E= \sum{_{s=10^{-5}}^2}{(|m_m,-m_e| + |\sigma_m- \sigma_e|)}$. A coarse parameter sweep was first performed, and the region that minimized the error (Fig. \ref{fig:fitsS1}, \ref{fig:fitsS2}) was subsequently swept using a finer resolution (Fig. \ref{fig:fitsS3}) to generate the final parameter sets shown in Fig. \ref{fig:fits}.

\subsection*{Sensitivity analysis}

To calculate the relative sensitivity,$S_{x,p} = (x_{pert,p} - x_0) /  x_0$ of each model variable, $x$, to each parameter, $p$, ~requires 1) solving Eqs. (\ref{mod}),(\ref{mod2}) to steady-state, $x_{0}$; 2) solving Eqs. (\ref{mod}),(\ref{mod2}) for each perturbed parameter to get the variable's perturbed value at steady-state, $x_{pert, p}$. For each value of $gal$ in $[10^{-5}, 0.0015, 10]$, each model parameter, $p$, was perturbed by $10\%$ of its original value.

\subsection*{Growth rate feedback analysis}

For the growth feedback (GFB) models, we rely on simple assumptions from  \cite{Bialek2012} to derive the growth rate law $r(S,G)$, given in Eq. (\ref{GrR}) (see Supplementary Materials). This was substituted into the $\gamma$ term of equations (\ref{mod}), (\ref{mod2}). Stochastic simulations using the Gillespie algorithm were carried out in Copasi \cite{Hoops2006}. Unless otherwise stated $V=45\mu m^3$, $R_{ext} = 0.0050 min.^{-1}$, $\rho=0.2$, and $K=15$ was assumed. The resultant trajectories were used to compute $P(G|S)$. The Gal1p steady state values with their respective error were inserted into $r(S,G)$ for different values of external galactose. In order to compute the mean growth rate, the probability distribution of Gal1p concentration obtained from the simulations was inserted into Eq. (\ref{fig:Mic}). To compute the optimal growth rate, $r_{opt}$, and the relative maximum growth rate accessible to the system, Eq. (\ref{optim}) is minimized in $P(G|S)$ to obtain the ideal $P(G|S)$ that in turn defines $r_{opt}$.

\section*{Results}

\subsection*{Fitting the galactose response of wild type and Gal80 promoter swapped strains}

\begin{figure*}[t]
 \centering
    \includegraphics[scale=0.85]{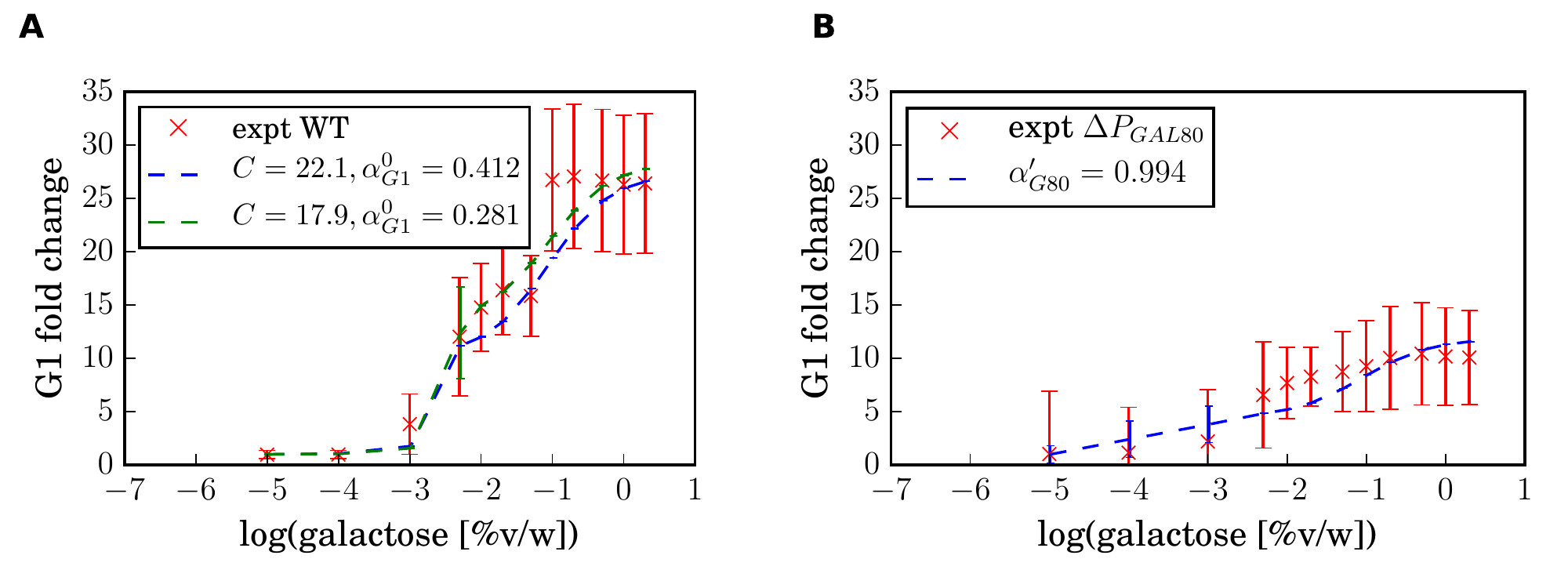}
    \caption{Galactose Gal1p fold change experimental data and model parameter fits. \textbf{A} Experimental data for the WT strain (red), the chosen parameter set (green), and the best fitting parameter set (blue). \textbf{B} Data for the $\Delta P_{GAL80}$ strain (red) and the best fitting parameter set (blue). All values represent mean fold changes, and error bars are standard deviations of the fold change.}
   \label{fig:fits}
\end{figure*}

Flow cytometry experiments for WT and $\Delta P_{GAL80}$ strains characterized fold changes in Gal1p-yCCFP as a function of galactose induction (Fig. \ref{fig:fits}A). For the WT strain galactose induction leads to a near 30-fold induction of Gal1p (Fig. \ref{fig:fits}A, red points), in agreement with separate measurements in similar conditions \cite{Ideker2001}. The distribution of G1 fold changes is broad for high galactose, but appears unimodal at all galactose levels (see \cite{Ferry2010} pp. 116-121). The parameter set ($C=17.9$ nM/min, $\alpha^0_{G1}=0.281$ nM/min) that minimized the error of the fit demonstrated bistability at $s=0.01\%$ (Fig. \ref{fig:fits}A, green error bar), but we found nearby parameters ($C=22.1$ nM/min, $\alpha^0_{G1}=0.412$ nM/min) that were monostable for all galactose values, and yet not very far off from the experimental data (Fig. \ref{fig:fits}A, blue curve, Fig. \ref{fig:fitsS1}). These models fits overestimate the saturation of Gal1p response at higher galactose concentration, partly due to an aberrantly low G1 fold change at $s=0.005\%$ in the WT experiment.

For the $\Delta P_{GAL80}$ experiment, the Gal1p response occurs at roughly the same galactose level, but the Gal1p fold change is noticeably reduced relative to WT (compare Fig. \ref{fig:fits}A,B). There is a broad range of Gal1p expression at low galactose concentrations (Fig. \ref{fig:fits}B, red error bars) that indicates bistability (see \cite{Ferry2010} p. 121). Using the parameter values found in the WT fit ($C=22.1$ nM/min, $\alpha^0_{G1}=0.412$ nM/min), we fit the $\Delta P_{GAL80}$ model to the experiment by varying the basal Gal80p production rate, and found that $\alpha^\prime_{G80}=0.994$ nM/min best reproduced the data. This parameter set recapitulated the decreased fold change in $\Delta P_{GAL80}$ relative to WT, and also produced bistability at lower galactose (Fig. \ref{fig:fits}B, blue curve and error bars).

\subsection*{Sensitivity analysis}
\begin{figure*}[t]
 \centering
    \includegraphics[scale=0.69]{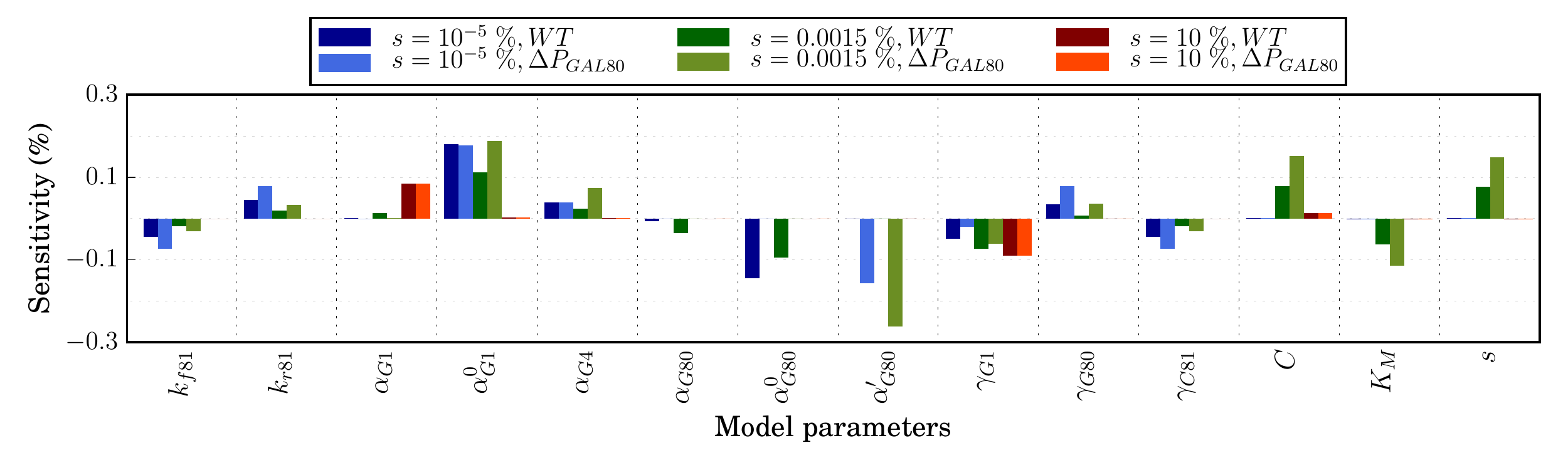}
    \caption{Sensitivity analysis for Gal1p in the WT and a $\Delta P_{GAL80}$ models. Only the parameters to which G1 is especially sensitive are shown. Different bar colors refer to different galactose values, $s$, while different color shades refer to different strains: WT and $\Delta P_{GAL80}$}
   \label{fig:sens_analysis}
\end{figure*}

 To characterize which parameters most affect model output we performed local sensitivity analysis for all variables of the two models at three galactose concentrations (Figures S1-S4). Fig. \ref{fig:sens_analysis} presents the results for G1, considering only the parameters to which it is most sensitive. The G1 steady state is particularly sensitive to parameter perturbations at $s=0.0015 \%$, because the system is strongly responding to galactose and less stable in transitioning from the OFF to the ON state. For $s=10 \%$, G1 is  more stable than for other $g$ levels, since G1 becomes sensitive only to its own production, $\alpha _{G1}$, and degradation, $\gamma_{G1}$, rates, whose impact on G1 increases with G1 and G4 concentrations, which in turn increase with galactose concentration. 
As expected, when the GAL80 promoter is removed, G1 is no longer sensitive to G80 parameters, except for $\alpha '_{G80}$ and its degradation rate. The strongest negative effect on G1 comes from the Gal80p production rate terms. For the $\Delta P_{GAL80}$  strain, G1 is only sensitive to $\alpha '_{G80}$, as $\alpha_{G80}$ and $\alpha^0_{G80}$ are not part of this model. While for the WT strain, G1 is only sensitive to $\alpha_{G80}$ and $\alpha^0_{G80}$ because $\alpha'_{G80}$ is not part of this model. G1 appears to be more sensitive to galactose concentration when $P_{GAL80}$ is removed than the WT strain, however, this is only because the unperturbed value of G1 at steady-state, $x_0$, is lower for the $\Delta P_{GAL80}$ strain than for WT, while the local difference $x_{pert,p}-x_0$ is comparable for both strains.
Finally, the degradation rates of Gal1p, Gal80p, and of the complex formed by these proteins, have a particularly strong effect on Gal1p concentration, with its sensitivity to $\gamma_{G1}$ being particularly high for high galactose levels, while its sensitivity to $\gamma_{G80}$ and $\gamma_{C81}$ is null at high galactose, as Gal80p's unbound concentration is depleted at that galactose level. Interestingly, for $s=10^{-5} \%$ Gal1p is barely sensitive to $\gamma_{G1}$ for the $\Delta P_{GAL80}$ strain, when compared to the WT strain. This is most likely because, for the $\Delta P_{GAL80}$, Gal80p is about $~3\times$ as high as in the WT strain, leading to a concentration of Gal1p $~3\times$ lower in the $\Delta P_{GAL80}$, and thus to a lower sensitivity to its own degradation rate.

\subsection*{Growth rate feedback}

To explore the influence of the cell growth rate on the galactose network, we first define how the growth rate is affected by sugar and the burden of expressing metabolic enzymes. Assuming the transport function from Eq. (\ref{eq:transp}) and a relation for how much energy from galactose can be used to make metabolic enzymes (see the Supplemental Materials), we derive the following contribution of the galactose network to the growth rate:

\begin{equation}\label{GrR}
\tilde{r}(S,G) = G(1-2\rho) + S +1 -\sqrt{(G-S)^2 + (1+G+S)}
\end{equation}

\noindent where $S$ is the external galactose, $G=G1$ is the amount of metabolic enzyme, and $\rho$ is the ratio between the cost of producing the enzyme and the maximum growth rate benefit achieved by burning galactose. 

In a stochastic nutrient information processing setting \cite{Bialek2012}, the mean growth rate can be defined as:

\begin{equation}\label{mGr}
 \langle \tilde{r} \rangle = \int dS P(S) \int dG P(G|S) \tilde{r}(G|S)
\end{equation}

\noindent where $P(S)$  represents probability of the cell sensing an amount of galactose $S$, and $P(G|S)$ the probability of G responding to S. The mutual information that flows between the external galactose concentration ($S$) and the relative concentration of Gal1p ($G$) is:

\begin{equation}\label{eq:MI}
I(G,S) = \int dG dS P(G,S) \log \frac{P(G,S)}{P(G)P(S)}
\end{equation}
where $P(G,S)$ is computed from the steady state values obtained from time traces of the Gillespie simulations of the models, and $P(S)$ is considered to be constant. $I(G,S)$ can be interpreted as a measure of the precision of the network in reading out the amount of sugar in the environment.

\begin{figure*}[t]
 \centering
    \includegraphics[scale=0.43]{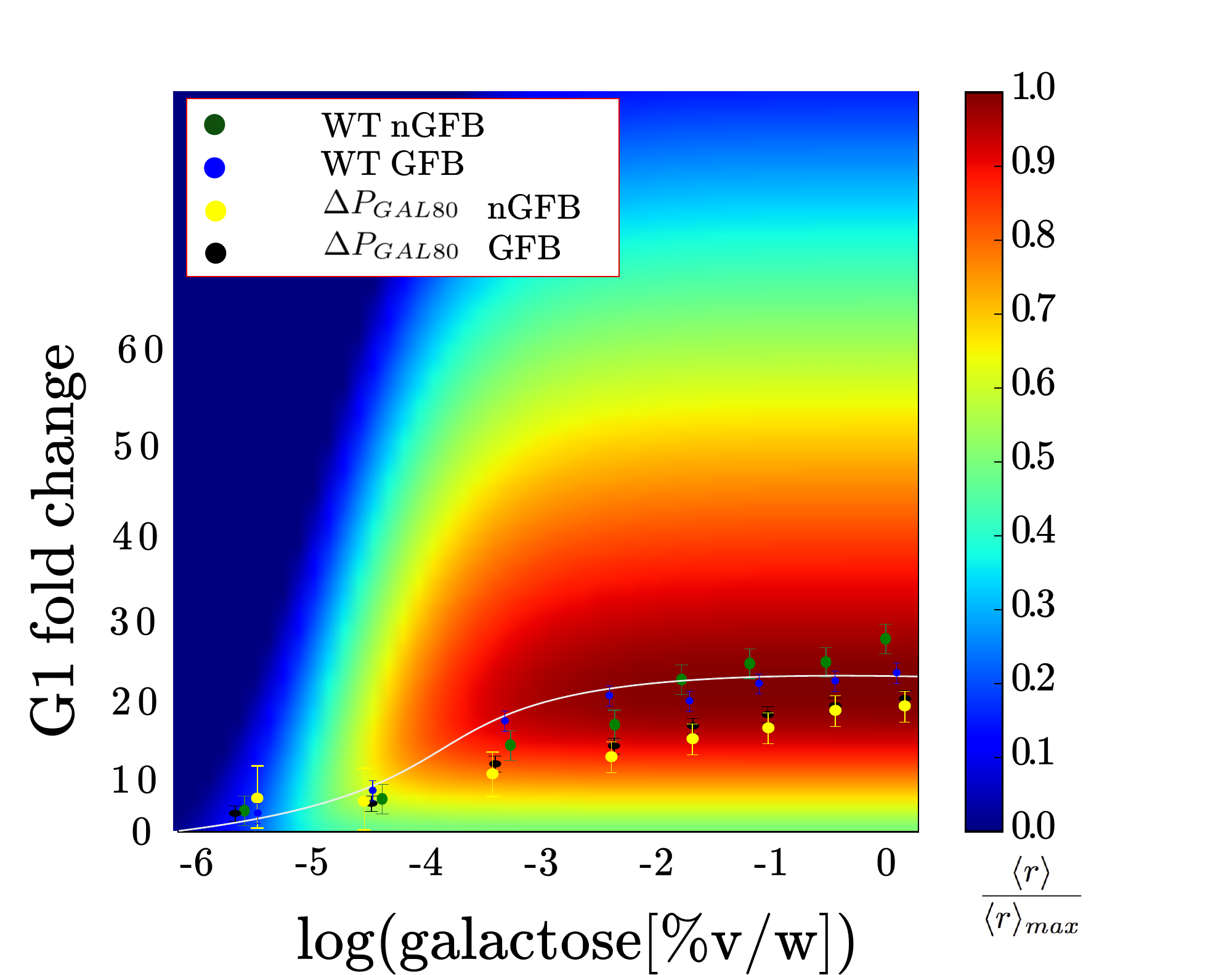}
     \includegraphics[scale=0.79]{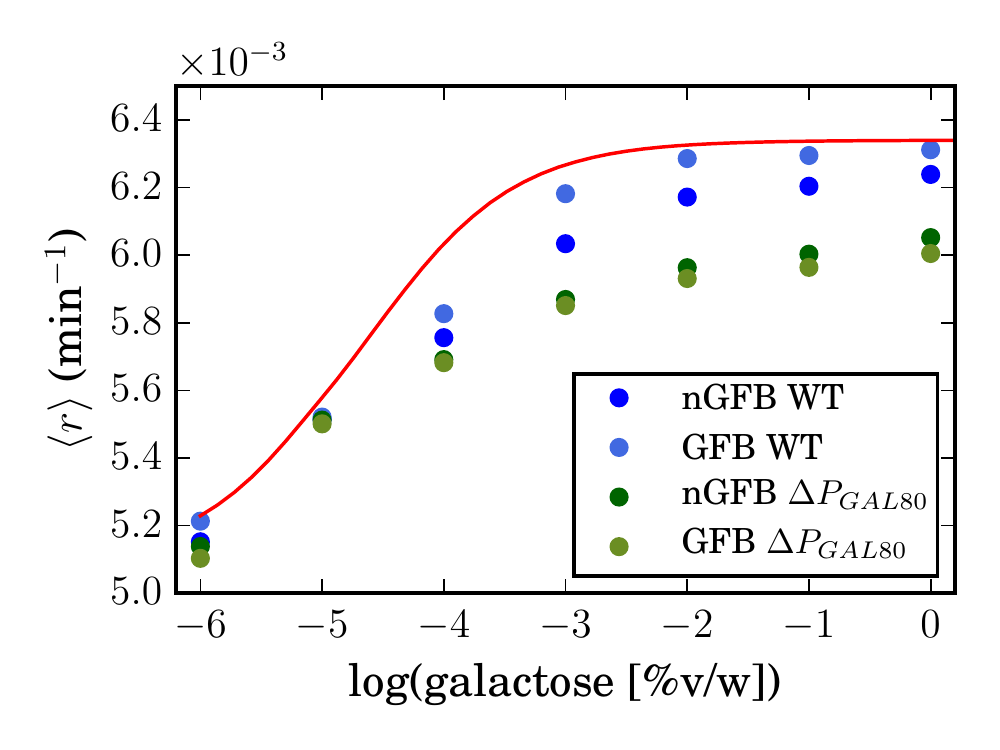}  \caption{\textbf{A}: heat map of the growth rate law, $r(S,G)$, and the steady state values of the Gal1p concentrations for different external galactose concentrations. The scale of the plot is relative to $\langle r\rangle_{max}$ that is the maximum of $r(S,G)$. The white line represents the optimal expression of $g$ given $s$ that maximizes the growth rate $r(s,g)$.   \textbf{B}: values of the mean growth rate $\langle r \rangle$ for the respective data plotted in panel \textbf{A}. The red line represents the maximum growth rate reachable, it is computed for the $s,g$ values that belong to the white line in panel \textbf{A}. }
   	\label{fig:Mic}
\end{figure*}

\begin{figure}[t]
 \centering
    \includegraphics[scale=0.58]{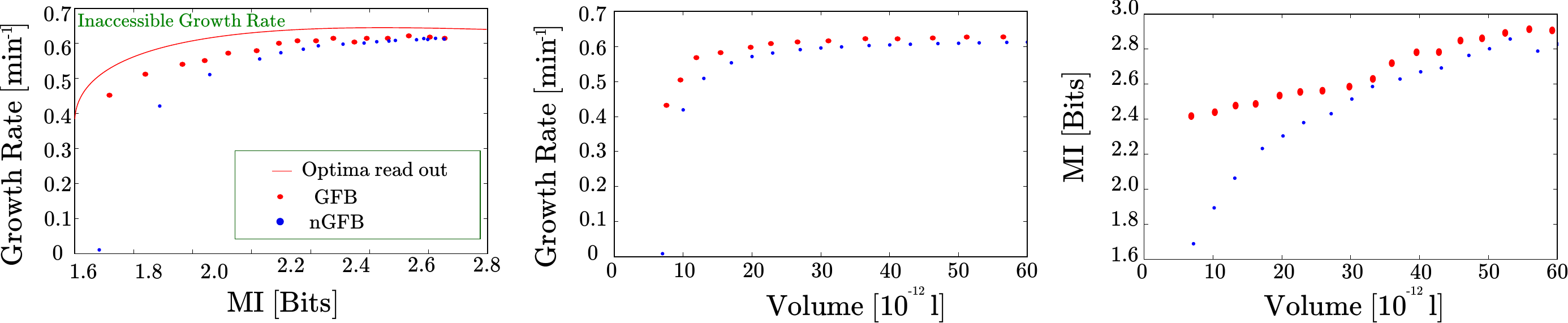}
    \caption{Relationship between growth rate, number of molecules and mutual information for the models with and without growth feedback. \textbf{A}: the average growth rate as a function of the mutual information between the external galactose concentration and the relative G1 expression (Eq. (\ref{eq:MI})). The average growth rate $\langle r \rangle$ has units of $10^{-3}min^{-1}$. The red line is the optimal growth rate, a solution to the optimization problem given by Eq. (\ref{optim}). \textbf{B,C}: The effect of cell volume on the average growth rate (\textbf{B}), and mutual information (\textbf{C}). }
   \label{fig:fig5}
\end{figure}

\begin{figure}[H]
 \centering    \includegraphics[scale=0.85]{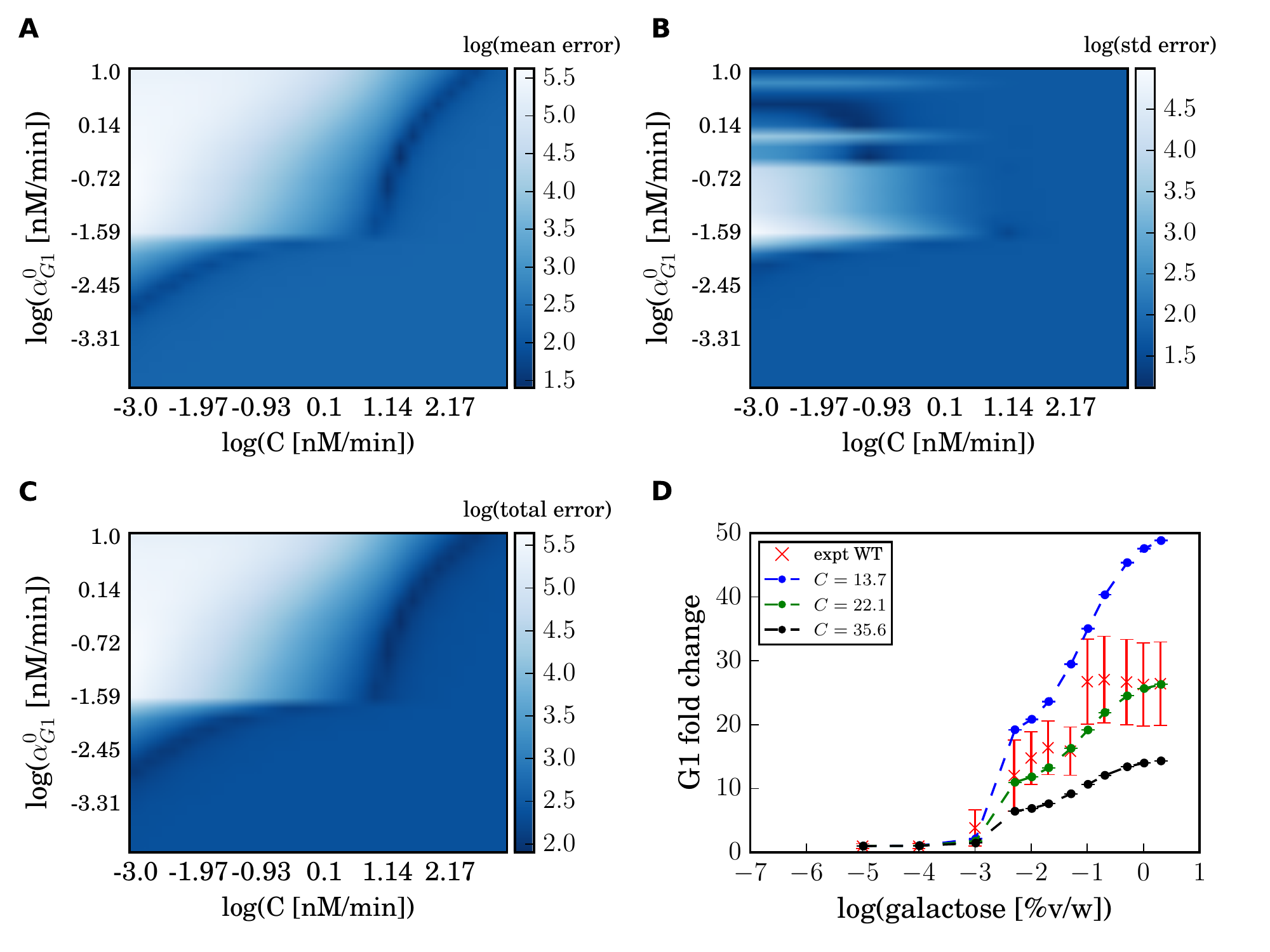}
    \caption{Coarse scan of parameter-error space for fitting the WT experiments. Parameter surfaces showing the \textbf{A} error mean, \textbf{B} error standard deviation, and \textbf{C} total error. \textbf{D} Optimal parameter set plotted with the WT experimental data, and solutions for nearby parameters}
   \label{fig:fitsS1}
\end{figure}

\begin{figure}[H]
 \centering
    \includegraphics[scale=0.85]{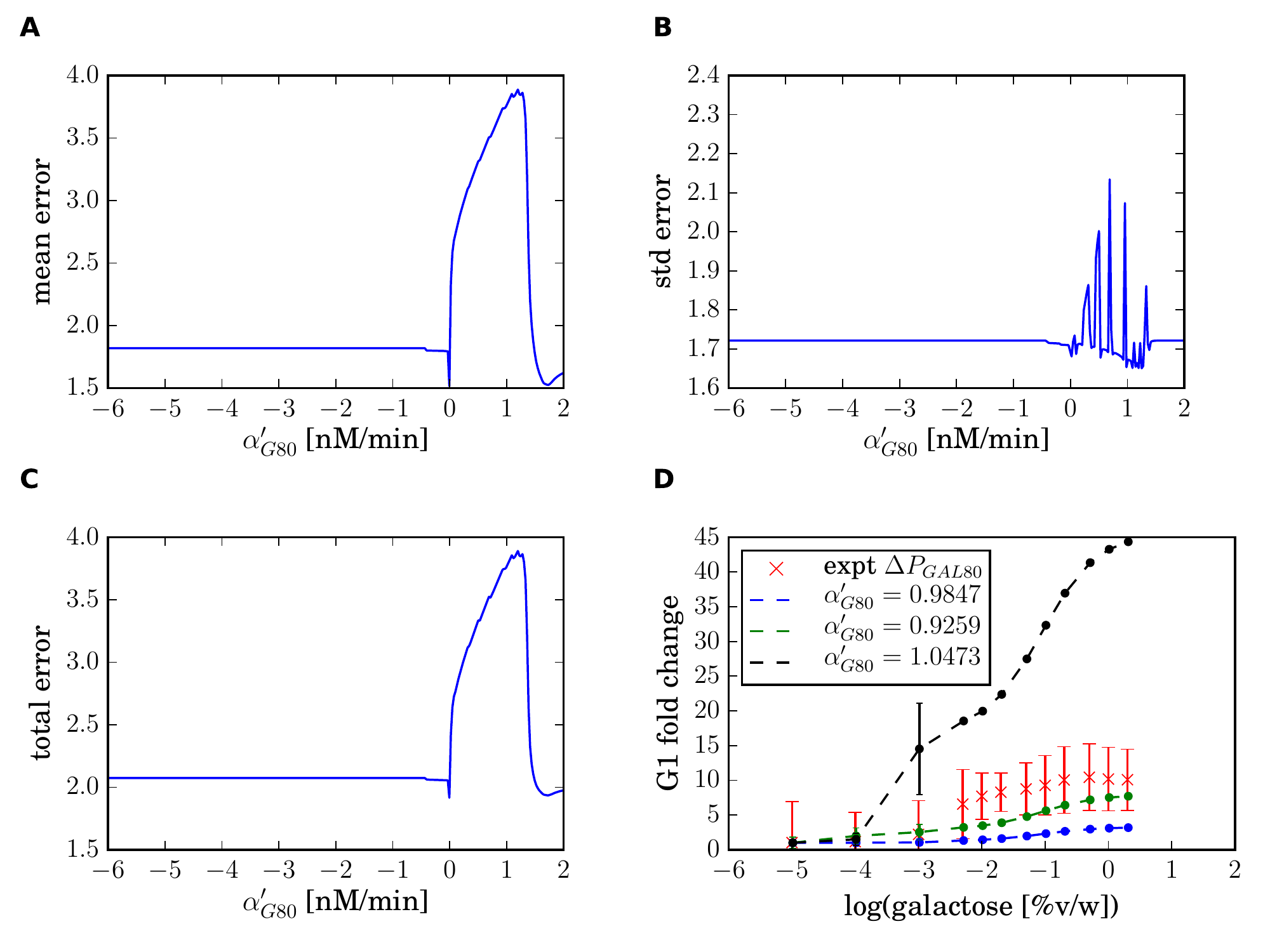}
    \caption{Coarse scan of parameter-error space for fitting the $\Delta P_{GAL80}$ experiments. Parameter surfaces showing the \textbf{A} error mean, \textbf{B} error standard deviation, and \textbf{C} total error. \textbf{D} Optimal parameter set solution plotted with the $\Delta P_{GAL80}$ experimental data, and solutions for nearby parameters}
   \label{fig:fitsS2}
\end{figure}

\begin{figure}[H]
 \centering
    \includegraphics[scale=0.85]{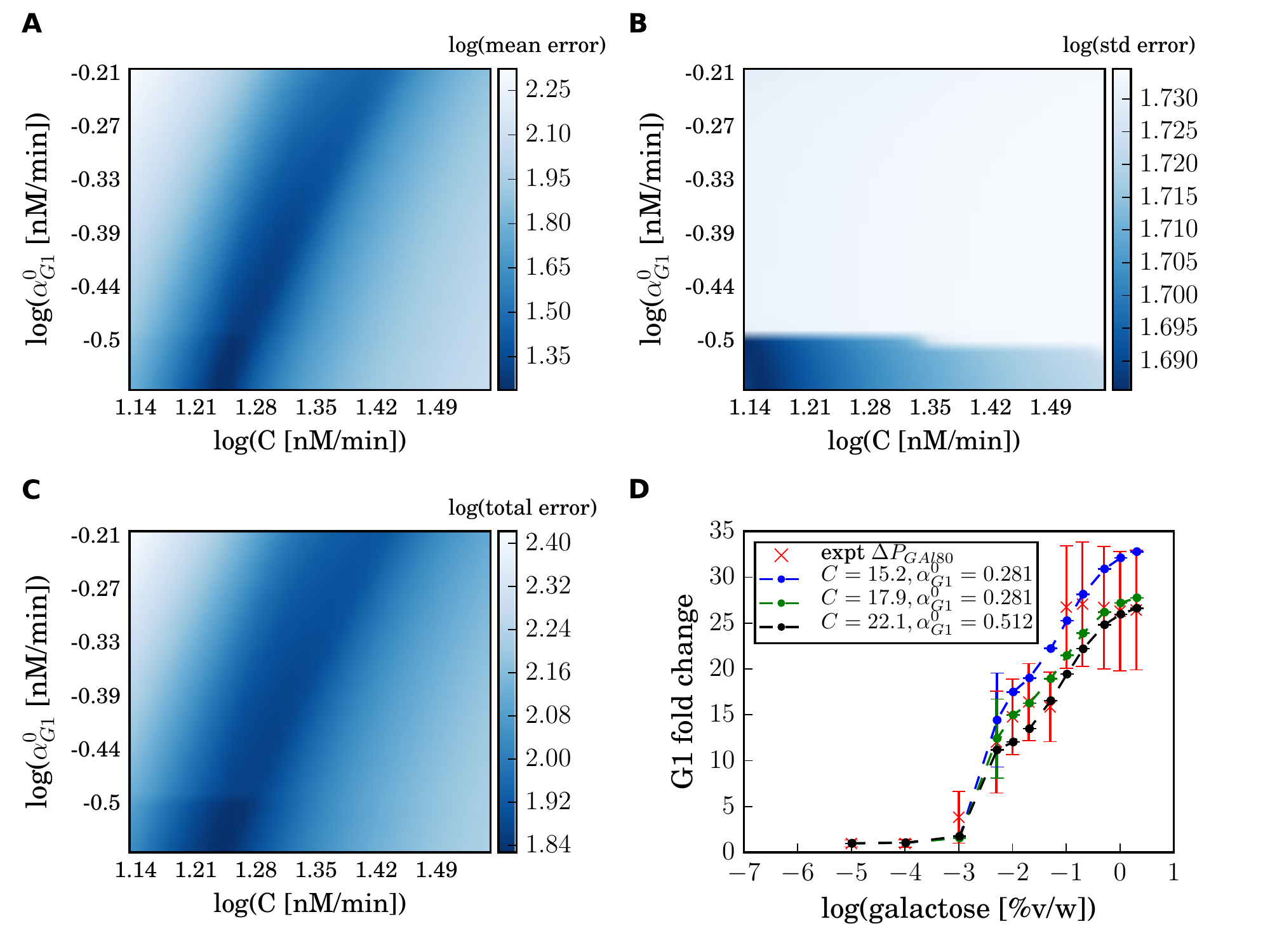}
    \caption{Finer scan of parameter-error space for fitting the WT experiments. Parameter surfaces showing the \textbf{A} error mean, \textbf{B} error standard deviation, and \textbf{C} total error. \textbf{D} Optimal parameter set plotted with the WT experimental data, and solutions for nearby parameters}
   \label{fig:fitsS3}
\end{figure}

To see how this formulation affects the galactose network, we substitute Eq. (\ref{GrR}) into each $\gamma$ term of equations (\ref{mod}),(\ref{mod2}). From the sensitivity analysis (Fig. \ref{fig:sens_analysis}), we know Gal1p will be sensitive to this change. Fig. \ref{fig:Mic}A shows the galactose, Gal1p fold change, and growth rate superimposed on the growth rate law Eq. (\ref{GrR}), assuming $\rho=0.2$, for the WT and $\Delta P_{GAL80}$, with the growth feedback (GFB) and without it (nGFB). For the $\Delta P_{GAL80}$ cases the growth feedback does not have much of an effect on Gal1p induction due to the bistable response in this model. For WT the GFB causes a very small decrease in G1 fold change for low galactose, with a further decrease at high galactose, reflecting the fact that the growth rate law causes an additional negative feedback for G1 in Eq. (\ref{mod}). Figure \ref{fig:Mic}B shows that the WT GFB model has higher growth rates than the other models at all galactose levels, as it is able to increase mutual information, $I(G,S)$, and stay closer to the optimal induction curve in Fig. \ref{fig:Mic}A (see Fig. \ref{fig:fig5}). 

Fig. \ref{fig:fig5} demonstrates that the WT GFB model performs better than nGFB for small volumes, $V$, or low number of molecules, $N$, while for high $V$ the two models perform similarly. In this context $V$ can be seen as a proxy for the noise. The volume, $V= 45 \mu m^3$ we use in the simulations are in the biological range for haploid {\emph S. cerevisiae} (15-70 $\mu m^3$ \cite{Jorgensen2007}). During the cell cycle the increasing volume of the cell changes the precision of the network and so the growth rate performance. Observing Fig. \ref{fig:fig5}B one can speculate that young yeast cells grow slower than cells that are close to division, because the processing of galactose is much more accurate for big volumes and high number of molecules. Indeed the key parameters to be tuned in order to optimize mutual information and the growth rate, are the noise strength and the ability to precisely regulate the expression of G1. From Fig \ref{fig:fig5}C, we can see that the feed back model is able to read out the environment more accurately by lowering the mutual information. Furthermore, introducing a dynamical interaction between gene expression and growth rate in the GFB model enables the system to better approach the optimal growth rate even at large $V$, although with diminishing returns.

\section*{Discussion and Conclusion}

Our modification of the Venturelli model \cite{Venturelli2012} enabled us to fit the large, monostable Gal1p response to galactose seen in the wild type strain. Venturelli found that knocking out the GAL1 and GAL3 feedbacks was sufficient to abolish bistability, but we show that it is not necessary: saturating activation by galactose and basal production of Gal1p can also achieve this effect. Nevertheless, fitting the wild type data shows that the model parameters ($C, \alpha_{G1}^0$) are actually quite close to the bistable regime (Fig. \ref{fig:fits}A). Acar {\emph et al} demonstrated that the transition from bistable to monostable responses can be induced by pre-growing cells in galactose even tens of hours prior to the induction from raffinose experiment \cite{Acar2005}. The fact that small changes in ($C, \alpha_{G1}^0$) can control this transition suggests that Gal1p and Gal2p (proportional to $C$) may be responsible for persistent memory of the ON state due to bistability \cite{Stockwell2015}, since they are both known to respond to galactose. It is also in accord with reports that Gal1p \cite{Venturelli2012} or Gal2p \cite{Hawkins2006} can mediate the transition from monostable to bistable responses.

The model could be further refined by testing this prediction in MFSC strains \cite{Ferry2010} with additional basal Gal1p production regulated by an inducible promoter. Our sensitivity analysis indicates this will have a significant effect at lower galactose concentrations. It also predicts a strong effect of Gal1p degradation in the ON state dominate the effects of dilution due to growth. As such, it would also be useful to extend the observations to other galactose inducing conditions (like galactose induction from glycerol) that would produce more distinguishable growth rates for model testing. 

Although our model is not the first model to incorporate growth into the galactose network \cite{Pannala2012}, it is the first to do so via the mechanism of protein dilution (also used in \cite{Furusawa2008}) and analysed in the context of optimizing information transmission through the galactose network \cite{Bialek2012}. Incorporating the growth rate feedback into the models improves optimal growth performance especially in the WT models since it can transmit information about the galactose with higher fidelity (no bistability). This analysis framework extends to other growth rate laws $r(S,G)$, including for instance multiple sugar sources \cite{Baumgartner2011, Pannala2012,Hermsen2015}, or effects from other proteins expressed and regulated indirectly by the sugar consumption \cite{Hui2015a}. In as much as all of our analysis applies only to steady state, a natural extension of the work would be to consider adapting it to account for growth in dynamic sugar conditions \cite{Bennett2008,Nguyen-Huu2015,Venturelli2015} that are very relevant for the competitive environments of natural yeast populations.

We hope the model and analysis presented here will inspire further development, and we thank the participants and organizers of the Ninth q-Bio Conference for enabling this collaboration.

\bibliographystyle{unsrt}
\bibliography{gal_network.bib}

\begin{thebibliography}{10}

\bibitem{Lavy2015}
Tali Lavy, Hayato Yanagida, and Dan~S Tawfik.
\newblock Gal3 binds gal80 tighter than gal1 indicating adaptive protein
  changes following duplication.
\newblock {\em Molecular biology and evolution}, page msv240, 2015.

\bibitem{Atauri2004}
P.~de~Atauri, D.~Orrell, S.~Ramsey, and H.~Bolouri.
\newblock Evolution of design principles in biochemical networks.
\newblock {\em Systems Biology}, 1:28--40(12), June 2004.

\bibitem{Acar2005}
Murat Acar, Attila Becskei, and Alexander van Oudenaarden.
\newblock Enhancement of cellular memory by reducing stochastic transitions.
\newblock {\em Nature}, 435(7039):228--232, 2005.

\bibitem{Orrell2006}
David Orrell, Stephen Ramsey, Marcello Marelli, Jennifer~J. Smith, Timothy~W.
  Petersen, Pedro~D. Atauri, John~D. Aitchison, and Hamid Bolouri.
\newblock Feedback control of stochastic noise in the yeast galactose
  utilization pathway.
\newblock {\em Physica D: Nonlinear Phenomena}, 217:64--76, 2006.

\bibitem{Ramsey2006}
Stephen~A Ramsey, Jennifer~J Smith, David Orrell, Marcello Marelli, Timothy~W
  Petersen, Pedro de~Atauri, Hamid Bolouri, and John~D Aitchison.
\newblock Dual feedback loops in the gal regulon suppress cellular
  heterogeneity in yeast.
\newblock {\em Nature genetics}, 38(9):1082--1087, 2006.

\bibitem{Bennett2008}
Matthew~R. Bennett, Wyming~Lee Pang, Natalie~a. Ostroff, Bridget~L.
  Baumgartner, Sujata Nayak, Lev~S. Tsimring, and Jeff Hasty.
\newblock {Metabolic gene regulation in a dynamically changing environment}.
\newblock {\em Nature}, 454(August):1119--1122, 2008.

\bibitem{Acar2010}
Murat Acar, Bernardo~F. Pando, Frances~H. Arnold, Michael~B. Elowitz, and
  Alexander van Oudenaarden.
\newblock {A General Mechanism for Network-Dosage Compensation in Gene
  Circuits}.
\newblock {\em Science}, 329(2010):1656--1660, 2010.

\bibitem{Venturelli2012}
Ophelia~S Venturelli, Hana El-Samad, and Richard~M Murray.
\newblock {Synergistic dual positive feedback loops established by molecular
  sequestration generate robust bimodal response.}
\newblock {\em Proceedings of the National Academy of Sciences of the United
  States of America}, 109(48):E3324--33, 2012.

\bibitem{cosentino2012}
Carlo Cosentino, Luca Salerno, Antonio Passanti, Alessio Merola, Declan~G
  Bates, and Francesco Amato.
\newblock Structural bistability of the gal regulatory network and
  characterization of its domains of attraction.
\newblock {\em Journal of Computational Biology}, 19(2):148--162, 2012.

\bibitem{Pannala2012}
VR~Pannala, SJ~Hazarika, PJ~Bhat, S~Bhartiya, and KV~Venkatesh.
\newblock Growth-related model of the gal system in saccharomyces cerevisiae
  predicts behaviour of several mutant strains.
\newblock {\em Systems Biology, IET}, 6(2):44--53, 2012.

\bibitem{Apostu2012}
Raluca Apostu and Michael~C Mackey.
\newblock Mathematical model of gal regulon dynamics in saccharomyces
  cerevisiae.
\newblock {\em Journal of theoretical biology}, 293:219--235, 2012.

\bibitem{Venturelli2015}
Ophelia~S Venturelli, Ignacio Zuleta, Richard~M Murray, and Hana El-Samad.
\newblock {Population diversification in a yeast metabolic program promotes
  anticipation of environmental shifts.}
\newblock {\em PLoS biology}, 13:e1002042, 2015.

\bibitem{Peng2015}
Weilin Peng, Ping Liu, Yuan Xue, and Murat Acar.
\newblock {Evolution of gene network activity by tuning the strength of
  negative-feedback regulation}.
\newblock {\em Nature Communications}, 6:6226, 2015.

\bibitem{Nguyen-Huu2015}
Truong~D. Nguyen-Huu, Chinmaya Gupta, Bo~Ma, William Ott, Kre\v{s}imir
  Josi\'{c}, and Matthew~R. Bennett.
\newblock {Timing and Variability of Galactose Metabolic Gene Activation Depend
  on the Rate of Environmental Change}.
\newblock {\em PLOS Computational Biology}, 11:e1004399, 2015.

\bibitem{Ferry2010}
Michael~Stephen Ferry.
\newblock {\em Synthetic biology in yeast : reconstructing the galactose
  network to probe the role of feedback induction in response to metabolic
  stimuli}.
\newblock PhD thesis, UC San Diego, 2010.

\bibitem{Barnett1997}
J.~A. Barnett.
\newblock Sugar utilization by saccharomyces cerevisiae.
\newblock In F.~K. Zimmermann and K.-D. Entian, editors, {\em Yeast Sugar
  Metabolism}. Technomic Publishing Company, Inc., Lancaster, PA, 1997.

\bibitem{Kasahara11022000}
Toshiko Kasahara and Michihiro Kasahara.
\newblock Three aromatic amino acid residues critical for galactose transport
  in yeast gal2 transporter.
\newblock {\em Journal of Biological Chemistry}, 275(6):4422--4428, 2000.

\bibitem{Raser2004}
Jonathan~M Raser and Erin~K O'Shea.
\newblock Control of stochasticity in eukaryotic gene expression.
\newblock {\em Science}, 304(5678):1811--1814, 2004.

\bibitem{Wach1996}
Achim Wach.
\newblock Pcr-synthesis of marker cassettes with long flanking homology regions
  for gene disruptions in s. cerevisiae.
\newblock {\em Yeast}, 12(3):259--265, 1996.

\bibitem{Bialek2012}
William Bialek.
\newblock {\em {Biophysics : Searching for Principles}}.
\newblock 2012.

\bibitem{Hoops2006}
Stefan Hoops, Sven Sahle, Ralph Gauges, Christine Lee, J{\"u}rgen Pahle,
  Natalia Simus, Mudita Singhal, Liang Xu, Pedro Mendes, and Ursula Kummer.
\newblock Copasi—a complex pathway simulator.
\newblock {\em Bioinformatics}, 22(24):3067--3074, 2006.

\bibitem{Ideker2001}
Trey Ideker, Vesteinn Thorsson, Jeffrey~A Ranish, Rowan Christmas, Jeremy
  Buhler, Jimmy~K Eng, Roger Bumgarner, David~R Goodlett, Ruedi Aebersold, and
  Leroy Hood.
\newblock Integrated genomic and proteomic analyses of a systematically
  perturbed metabolic network.
\newblock {\em Science}, 292(5518):929--934, 2001.

\bibitem{Jorgensen2007}
Paul Jorgensen, Nicholas~P Edgington, Brandt~L Schneider, Ivan Rupe{\v{s}},
  Mike Tyers, and Bruce Futcher.
\newblock The size of the nucleus increases as yeast cells grow.
\newblock {\em Molecular biology of the cell}, 18(9):3523--3532, 2007.

\bibitem{Stockwell2015}
Sarah~R. Stockwell, Christian~R. Landry, and Scott~a. Rifkin.
\newblock {The yeast galactose network as a quantitative model for cellular
  memory}.
\newblock {\em Mol. BioSyst.}, 11:28--37, 2015.

\bibitem{Hawkins2006}
Kristy~M Hawkins and Christina~D Smolke.
\newblock The regulatory roles of the galactose permease and kinase in the
  induction response of the gal network in saccharomyces cerevisiae.
\newblock {\em Journal of Biological Chemistry}, 281(19):13485--13492, 2006.

\bibitem{Furusawa2008}
Chikara Furusawa and Kunihiko Kaneko.
\newblock A generic mechanism for adaptive growth rate regulation.
\newblock {\em PLoS Computational Biology}, 4(1):e3, 2008.

\bibitem{Baumgartner2011}
B.~L. Baumgartner, M.~R. Bennett, M.~Ferry, T.~L. Johnson, L.~S. Tsimring, and
  J.~Hasty.
\newblock {Antagonistic gene transcripts regulate adaptation to new growth
  environments}.
\newblock {\em Proceedings of the National Academy of Sciences},
  108:21087--21092, 2011.

\bibitem{Hermsen2015}
Rutger Hermsen, Hiroyuki Okano, Conghui You, Nicole Werner, and Terence Hwa.
\newblock A growth-rate composition formula for the growth of e. coli on
  co-utilized carbon substrates.
\newblock {\em Molecular systems biology}, 11(4):801, 2015.

\bibitem{Hui2015a}
Sheng Hui, Josh~M Silverman, Stephen~S Chen, David~W Erickson, Markus Basan,
  Jilong Wang, Terence Hwa, and James~R Williamson.
\newblock {Quantitative proteomic analysis reveals a simple strategy of global
  resource allocation in bacteria}.
\newblock {\em Molecular systems biology}, 11(784):1--15, 2015.

\end{thebibliography}

\newpage{}

\section*{Supplemental materials}

\subsection*{Model parameters}

\begin{center}
\begin{table}[H]
\begin{centering}
\begin{tabular}{|c|l|c|c|}
\hline 
Parameter & Description & Value & Units\tabularnewline
\hline 
\hline 
$k_{f81}$ & Forward binding rate of Gal1p to Gal80p & 100 & $(nM\cdot min)^{-1}$\tabularnewline
\hline 
$k_{r81}$ & Unbinding rate of Gal1p to Gal80p & 1500 & $min^{-1}$\tabularnewline
\hline 
$k_{f83}$ & Forward binding rate of Gal3p to Gal80p & 100 & $(nM\cdot min)^{-1}$\tabularnewline
\hline 
$k_{r83}$ & Unbinding rate of Gal3p to Gal80p & 1 & $min^{-1}$\tabularnewline
\hline 
$k_{f84}$ & Forward binding rate of Gal4p to Gal80p & 100 & $(nM\cdot min)^{-1}$\tabularnewline
\hline 
$k_{r84}$ & Unbinding rate of Gal4p to Gal80p & 25 & $min^{-1}$\tabularnewline
\hline 
$\alpha_{G1}$ & Gal1p production rate & 15 & $(nM\cdot min)^{-1}$\tabularnewline
\hline 
$\alpha_{G1}^{0}$ & Basal Gal1p production rate & 0.418 & $(nM\cdot min)^{-1}$\tabularnewline
\hline 
$\alpha_{G3}$ & Gal3p production rate & 0.9 & $(nM\cdot min)^{-1}$\tabularnewline
\hline 
$\alpha_{G4}$ & Gal4p production rate & 0.2 & $(nM\cdot min)^{-1}$\tabularnewline
\hline 
$\alpha_{G80}^{0}$ & Basal Gal80p production rate & 0.6 & $(nM\cdot min)^{-1}$\tabularnewline
\hline 
$\alpha'_{G80}$ & Constant Gal80p production rate & 0.994 & $(nM\cdot min)^{-1}$\tabularnewline
\hline 
$\alpha_{G80}$ & Gal80p production rate & 0.9 & $(nM\cdot min)^{-1}$\tabularnewline
\hline 
$K_{G1}$ & GAL1 transcriptional feedback threshold & 8 & $nM$\tabularnewline
\hline 
$K_{G3}$ & GAL3 transcriptional feedback threshold & 8 & $nM$\tabularnewline
\hline 
$K_{G80}$ & GAL80 transcriptional feedback threshold & 2 & $nM$\tabularnewline
\hline 
$n_{1}$ & GAL1 Hill coefficient & 3 & Dimensionless\tabularnewline
\hline 
$n_{3}$ & GAL3 Hill coefficient & 2 & Dimensionless\tabularnewline
\hline 
$n_{80}$ & GAL80 Hill coefficient & 2 & Dimensionless\tabularnewline
\hline 
$\gamma_{G1}$ & Gal1p degradation rate & 0.004 & $min^{-1}$\tabularnewline
\hline 
$\gamma_{G3}$ & Gal3p degradation rate & 0.004 & $min^{-1}$\tabularnewline
\hline 
$\gamma_{G4}$ & Gal4p degradation rate & 0.004 & $min^{-1}$\tabularnewline
\hline 
$\gamma_{G80}$ & Gal80p degradation rate & 0.004 & $min^{-1}$\tabularnewline
\hline 
$\gamma_{C81}$ & Gal1p-Gal80p (C81) degradation rate & 0.004 & $min^{-1}$\tabularnewline
\hline 
$\gamma_{C83}$ & Gal3p-Gal80p (C83) degradation rate & 0.004 & $min^{-1}$\tabularnewline
\hline 
$\gamma_{C84}$ & Gal4p-Gal80p (C84) degradation rate & 0.004 & $min^{-1}$\tabularnewline
\hline 
$\epsilon$ & Scaling factor & 0.1 & Dimensionless\tabularnewline
\hline 
$C$ & Maximal galactose transport rate & 22.1 & $(nM\cdot min)^{-1}$\tabularnewline
\hline 
$K_{M}$ & Michaelis-Menten equilibrium constant of the facilitated diffusion
reaction & 0.086 & $w/v\%$\tabularnewline
\hline 
\end{tabular}
\par\end{centering}

\caption{Parameters used on the model, both for the WT and the $\Delta P_{GAL80}$ strain. $\alpha^0_{G80}$ and $\alpha_{G80}$ are only used in the WT model, while $\alpha'_{G80}$ is only used in the model where $P_{GAL80}$ is removed.}
\label{tab1}
\end{table}
\par\end{center}

\subsection*{Analysis scripts}

Analysis scripts for fitting, sensitivity analysis, and growth feedback are attached in Supplementary Data Files or available $\href{<https://github.com/BarttraB/GAL_Model_scripts_qbio2015>}{here}$.

\subsection*{Fitting}

\subsection*{Sensitivity analysis}

\begin{figure}[H]
 \centering
    \includegraphics[scale=0.7]{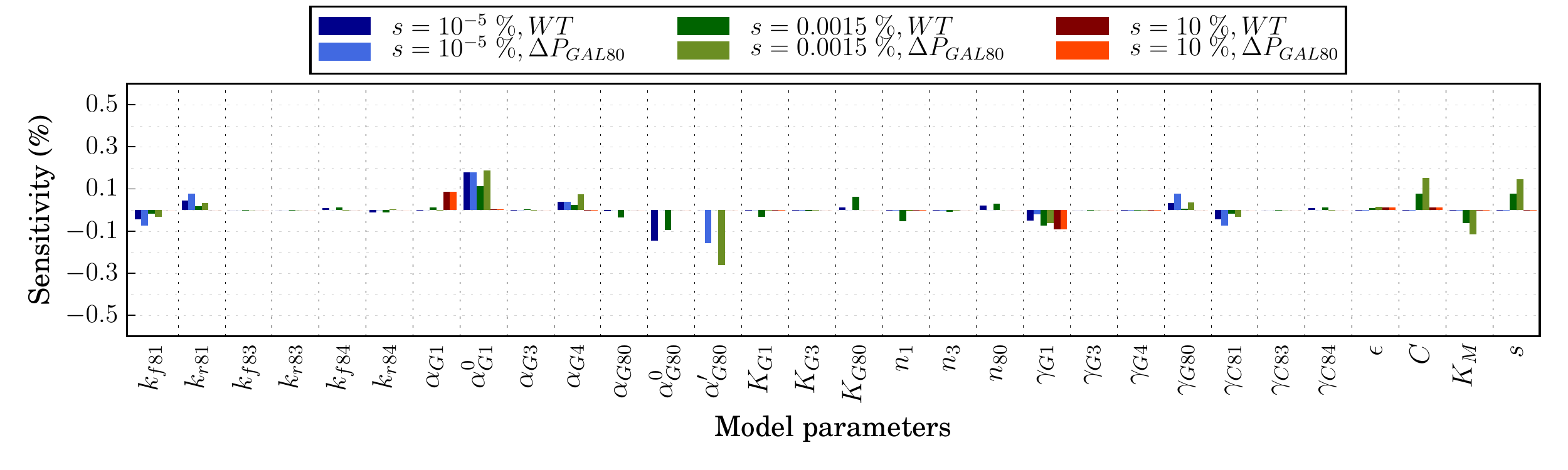}
    \caption{Gal1p steady state sensitivity}
   \label{fig:sens1}
\end{figure}

\begin{figure}[H]
 \centering
    \includegraphics[scale=0.7]{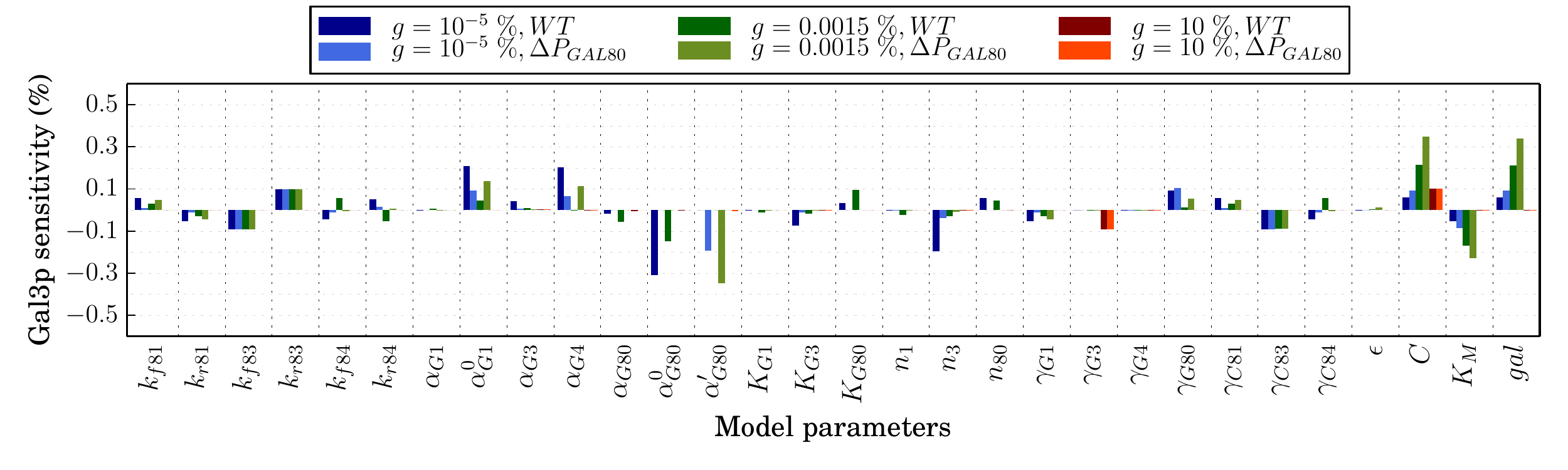}
    \caption{Gal3p steady state sensitivity}
   \label{fig:sens2}
\end{figure}

\begin{figure}[H]
 \centering
    \includegraphics[scale=0.7]{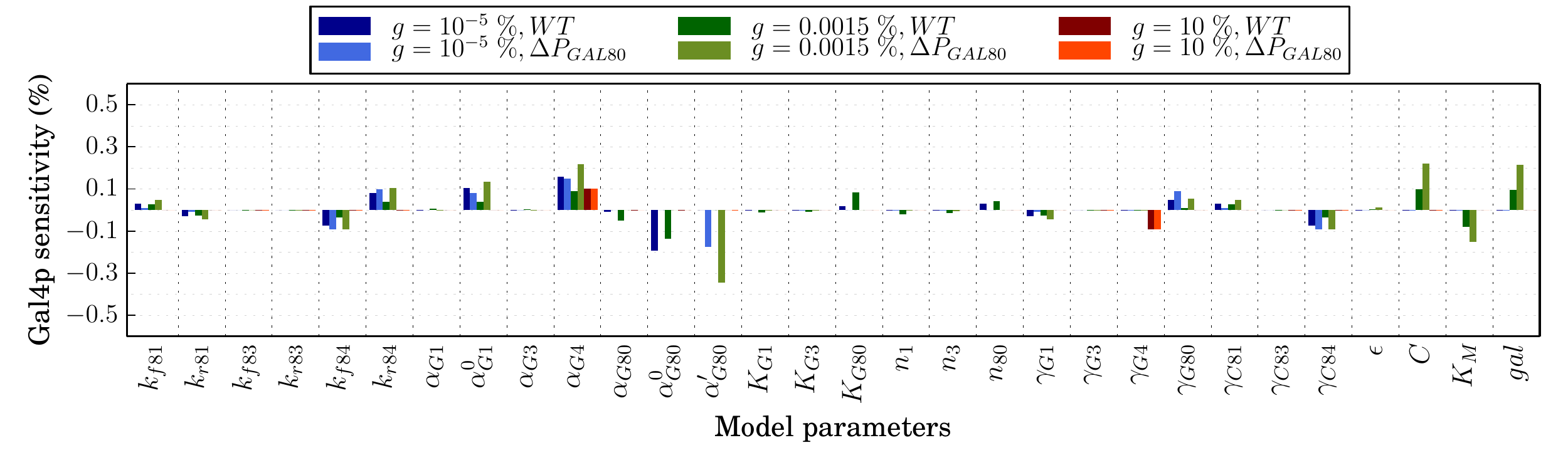}
    \caption{Gal4p steady state sensitivity}
   \label{fig:sens3}
\end{figure}

\begin{figure}[H]
 \centering
    \includegraphics[scale=0.7]{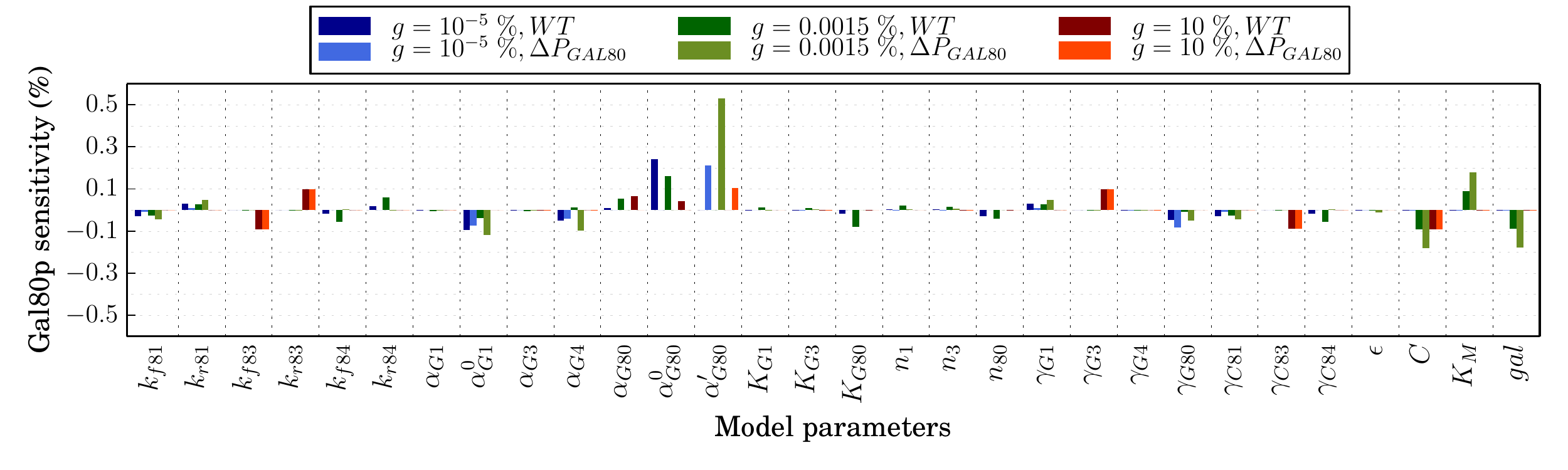}
    \caption{Gal80p steady state sensitivity}
   \label{fig:sens4}
\end{figure}

\subsection*{Growth feedback derivation and analysis}
Here we derive the growth rate law from equation \ref{GrR}. This term defines the contribution of the galactose metabolism to the overall cell growth rate. The cell growth rate takes into account a part relative to the Galactose metabolism ($r(s,g)$) plus an external function, $R_{ext}$, which sums up other processes that contribute to cell growth. $R_{ext}$ can  be considered constant.
To use the currently available substrates, the cell has to express the relevant enzymes involved in metabolizing those substrates. In this context it is possible to assume there is one  substrate at concentration $s$ (external galactose) and one relevant metabolic enzyme at expression level $g$ ($G1$ is the kick stater of the glycolisis of galactose). The cell will then grow at some rate $r(s,g)$, which depends both on the state of the environment ($s$) and on the cell's internal state ($g$). We can assume that the growth rate of the cell is a tradeoff between two effects: 1) growth depends $G1$ expression, so that the available substrate (galactose) can be metabolized - thus growth should be faster if there is either more substrate or more enzyme; 2) producing the enzyme itself consumes resources, which will slow the growth \cite{Hui2015a}. In the limit of low substrate concentrations, this cost can become dominant, and growth would stop if the cell tried to produce too many enzymes.  The goal of the network that reads out the substrate is to fine regulate the amount of expression of $g$ in order to have the right number of enzymes necessary to consume the amount of substrate available $s$. 

At steady state the influence of the galactose network on the growth rate can be expressed as:
\begin{equation}\label{sys}
\begin{aligned}
s_{in} &= C\frac{s}{s+K_M} \\
s_a &= s_{in} - \xi g \frac{s_a}{s_a+K}= C \frac{s}{s+K_M} - \xi g \frac{s_a}{s_a+K}\\
r(s_a,g) &= \xi g \frac{s_a}{s_a+K} - \mu g
\end{aligned}
\end{equation}
where $s$ is the amount of external galactose, $s_{in}$ is the amount of nutrients inside the cell, $s_a$  quantifies the amount of sugar  in the cell available for growing. The total amount of galactose available in the cell at the steady state is given by the amount of sugar it gets from the environment minus the amount of sugar that is consumed by $g$.

Solving eq. (\ref{sys}) for $s_a = f(s_{in})$ one can directly express the growth rate as a function of the external galacotse concentration $s$:
\begin{equation}\label{Gr}
r(s_{in},g) = g(\xi/2 -\mu) + s_{in}/2 +\xi K -\frac{\xi}{2}\sqrt{(k+g-s_{in})^2 + 4K s_{in}}
\end{equation}

Rescaling the system to:
\begin{equation}
\begin{aligned}
\frac{s_{in}}{2K} \rightarrow S \\
\frac{g}{2K}\rightarrow G
\end{aligned}
\end{equation}
we get the rescaled growth rate law in units of $\xi K$:
\begin{equation}
\tilde{r}(S,G) = G(1-2\rho) + S +1 -\sqrt{(G-S)^2 + (1+G+S)}
\end{equation}
where now the only parameter is $\rho = \frac{ \mu }{\xi}$, which represents the ratio between the cost of making the enzyme and the growth rate advantage from burning galactose.

Since the overall growth rate of the cell must also take into account the contribution other mechanisms that contribute to growth, $R_{ext}$, which we consider to be constant. We set $R_{ext} = 0.0050$ to account for the doubling time shift from 110 mins. in raffinose to 120 mins. in galactose \cite{Ferry2010} and $K = 15$ to account for the G1 fold change data.

The mean growth rate is defined as:
\begin{equation}
 \langle \tilde{r} \rangle = \int dS P(S) \int dG P(G|S) \tilde{r}(G|S)
\end{equation}
where $P(S)$  represents how the input signal fluctuates, or in other 
words, what is the probability distribution such that the cell gets $s$ 
level of galactose. 

Based on an idea from Bialek \cite{Bialek2012}, we can ask: how precisely must $G$ track $S$ in order to grow at a certain rate, $ \langle r \rangle$? This optimization problem can be solved using the Lagrange multipliers to define a function:

\begin{equation}\label{optim}
\mathcal{F}[ P(G|S) ] = I(G,S) - \lambda \langle r \rangle - \int \mu(S) dS \int dG P(G|S)
\end{equation}
\noindent where $\lambda$ and $\mu(S)$ are the Lagrange multipliers. To get the optimal read out system we need to minimize it with respect to $P(G|S)$. For a given $\langle r \rangle$, this defines the optimal read out system $P(G|S)$ such that the growth rate is maximized. Or rather for given $I(g,s)$ it is possible to define the maximum growth rate accessible $r_{opt}$.

We can use this tool to compare models where the dilution rate $\gamma$ is considered constant and where it changes dynamically with the growth rate. Changing the number of molecules in the cell, i.e., the amount of intrinsic noise of the network, we want to see how these two models perform in terms of growth rate. Indeed, we can use the parameter $N$ as a proxy for the intrinsic noise of the network: when $N$ is small the noise is high and vice-versa.

What we expect is that the negative feed back of G1 on its own expression keeps the noise low also when the initial number of molecules in the cell is small. When $\gamma$ is constant (nGFB) the noise in the expression of G1 is large when $N$ is low and decreases for increasing $N$. Note that both models tend to the optimal at the high limit of $N$.  Our aim is to compute how the gal network reads out the environment and compare it with the optimal case.

\end{document}